\documentclass{ws-book9x6}
\usepackage{mathrsfs}
\usepackage{amsfonts}
\usepackage{bm}
\usepackage{ws-book-thm}           % comment this line when `amsthm` is used
\usepackage{ws-book-har}           % citation - Author-Date system
\usepackage[pdfpagelabels=false]{hyperref}  % \label, \ref and \cite are recommended
\newcommand{\bsub}{\begin{subequations}}
\newcommand{\esub}{\end{subequations}}
\newcommand{\beq}{\vspace{0.5em}\begin{equation}}
\newcommand{\eeq}{\end{equation}\vspace{0.5em}}
\newcommand{\beqn}{\vspace{0.5em}\begin{eqnarray}}
\newcommand{\eeqn}{\end{eqnarray}\par\vspace{0.5em}\noindent}
\newcommand{\br}{{\mathbf{r}}}

\renewcommand{\vec}[1]{\mbox{\boldmath $#1$}}

\begin{document}

\setcounter{chapter}{6}
\chapter{Structure of hypernuclei in relativistic approaches\label{ch7}}

\author{K. Hagino$^{1,2}$ and J. M. Yao$^{1,3}$}
\address{$^1$Department of Physics, Tohoku University, Sendai 980-8578, Japan \\
$^2$Research Center for Electron Photon Science, Tohoku University, 1-2-1 Mikamine, Sendai 982-0826, Japan \\
$^3$ School of Physical Science and Technology, Southwest University, Chongqing 400715, China}

\section{Introduction}

Hypernuclei consist of neutrons, protons, and one or more hyperons such as $\Lambda$, $\Sigma$,
and $\Xi$. In Table \ref{table1}, we summarize properties of those hyperons.
The most extensively studied hypernuclei are single-$\Lambda$ hypernuclei, which contain one $\Lambda$ particle inside ordinary nuclei. 
Since the first discovery of $\Lambda$ hypernucleus by Danysz and Pniewski in 1953~\cite{DP53}
(see also Ref. ~\cite{W04}), more than thirty $\Lambda$ hypernuclei ranging from $^3_\Lambda$H up to $^{208}_\Lambda$Pb have been produced. Several properties of hypernuclei, such as
the mass number dependence of $\Lambda$ binding energy, spin-orbit splittings, and electromagnetic transitions,
have been revealed using reaction spectroscopy with ($K^-, \pi^-$), ($\pi^+, K^+$), and ($e, e'K^+$)
reactions as well as $\gamma$-ray spectroscopy~\cite{HT06}. Decay properties of hypernuclei have also been studied ~\cite{AG02,BMZ90}.

An important motivation to study hypernuclei is to extract information on 
baryon-baryon interactions including the strangeness degrees of freedom.
Such information is crucial in order to understand neutron stars, in which hyperons may emerge
in the inner part ~\cite{V13,SB08},
since the first hyperon to appear may be different depending on the properties of baryon-baryon interaction. Because hyperon-nucleon and hyperon-hyperon scattering experiments are difficult to perform, the structure of hypernuclei has been playing a
vital role in order to shed light on baryon-baryon interactions. 

From the nuclear structure point of view, an important feature of $\Lambda$ particle is that it carries a strangeness degree of freedom and therefore it does not suffer from the Pauli principle from other nucleons and can thus deeply penetrate into the
nuclear interior. This may induce a change of {\it e.g.,} nuclear size~\cite{MBI83,HKMM99}, the
density distribution~\cite{HKYMR10},
deformation properties~\cite{Z80,ZSSWZ07,WH08,SWHS10,WHK11,YLHWZM11,LZZ11,IKDO11,Isaka12,Isaka14},
collective excitations~\cite{YLHWZM11,HYM13,MH12,MHYM14,Isaka13},
the neutron drip-line~\cite{VPLR98,ZPSV08},
and fission barrier~\cite{MCH09,MC11}. This property is referred to as the impurity effect of $\Lambda$ particle, which has
also been one of the most important topics in hypernuclear physics. Moreover, some new states called {\em supersymmetric states}~\cite{Dalitz76,Zhang82,Motoba85} that cannot exist in ordinary nuclei can appear in hypernuclei, the study of which greatly 
broadens our knowledge on nuclear physics.

\begin{table}[b]
%\centering
%\tabcolsep=6pt
 \tbl{Properties of nucleons and hyperons from the Particle Data Group~\cite{PDG2012}. }
{\begin{tabular}{ccccccc}
  \hline\hline
particle & spin and parity & charge & mass (MeV/c$^2$) & mean-life  (sec.) & quark content \\
\hline
p & 1/2$^+$ & +$e$ & 938.27 & $>2.1\times10^{29}$ & uud \\
n & 1/2$^+$ & 0 & 939.57 & 880 & udd \\
\hline
$\Lambda$ & 1/2$^+$ & 0 & 1115.68 & $2.63(2)\times10^{-10}$ & uds \\
$\Sigma^+$ & 1/2$^+$ & +$e$ & 1189.37& $0.80 \times10^{-10}$ & uus \\
$\Sigma^0$ & 1/2$^+$ & 0 & 1192.64 & $7.4(7) \times10^{-20}$ & uds \\
$\Sigma^-$ & 1/2$^+$ & $-e$ & 1197.45 & $1.48(1) \times10^{-10}$ & dds \\
$\Xi^0$ & 1/2$^+$ & 0 & 1314.86 & $2.90(9) \times10^{-10}$ & uss \\
$\Xi^-$ & 1/2$^+$ & $-e$ & 1321.71 & $1.64(1) \times10^{-10}$  & dss \\
\hline \hline
\end{tabular}}
\label{table1}
\end{table}

Theoretically, the structure of hypernuclei
has been studied with
the cluster model~\cite{MBI83,HKMM99,BMZ90,Motoba85},
the shell model ~\cite{Dalitz78,Gal71,Millener,Millener13},
the mean-field approaches
~\cite{Rayet76,Rayet81,YBZ88,ZSSWZ07,WHK11,BW77,BB81,MZ89,VPLR98,ST94,WH08,LZZ11},
and the antisymmetrized molecular dynamics (AMD) ~\cite{IKDO11,Isaka12,Isaka13,Isaka14}.
Recently, an ab-initio method has also been
applied to hypernuclei~\cite{abinitio}.
Among these theoretical methods, the mean-field approach has an advantage in that it can
be globaly applied from light to heavy hypernuclei. The mean-field approach can also be
directly applied to neutron stars, in which
hyperons play an important role.
Both non-relativistic Skyrme-Hartree-Fock ~\cite{Rayet76,Rayet81}
and relativistic mean-field method ~\cite{BW77,BB81,MZ89} have been applied
to hypernuclear physics.

In this Chapter, we review the relativistic mean-field approach to various aspects of
hypernuclei.
It was Brockmann and Weise who first applied this
approach to hypernuclei ~\cite{BW77}.
At that time, it had been already
observed experimentally that the spin-orbit splittings in hypernuclei
are significantly smaller than that in ordinary nuclei ~\cite{BFK78}.
The relativistic approach was suitable for a discussion of spin-orbit splittings in hypernuclei,
as the spin-orbit interaction is naturally emerged with the relativistic framework.
Since this seminal work, many developments have been done and the relativistic approach
has been applied both to the ground state and to excited states of hypernuclei.
We summarize those developments in the Chapter.
We shall concentrate on $\Lambda$-hypernuclei and do not discuss $\Sigma$ hypernuclei~\cite{B81,Boussy81,ZQ85,MJ94,Glendenning93},
since the theoretical treatment is rather similar
to each other.

The Chapter is organized as follows.
In Sec. 2, we detail how the relativistic Lagrangian can be extended to hypernuclei.
We will discuss both the meson exchange and the point coupling approaches.
In Sec. 3, we discuss the ground state properties of hypernuclei. These include the binding
energy of $\Lambda$ particle, the spin-orbit splittings, the hypernucler magnetic
moments, an extention of neutron drip-line, and the deformation properties.
In Sec. 4, we discuss how collective excitations are altered in hypernuclei, employing the beyond mean-field approach. In Sec. 5,  we focus our attention on anti-$\Lambda$ hypernuclei. We will
discuss the spin symmetry and the tensor coupling effects. In Sec. 6, we will briefly discuss
multi-$\Lambda$ hypernuclei as well as neutron stars. We will then summarize this Chapter in
Sec. 7.

\section{Relativistic mean-field approaches to hypernuclei}

\subsection{Lagrangian density with hyperons}
 In this section, we will introduce the general framework of RMF approaches to hypernuclei.
The starting point of the RMF approaches for hypernuclei is the following Lagrangian density
  \begin{equation}
  \label{Lag1}
  {\cal L} = {\cal L}^{\rm free} + {\cal L}^{\rm em} + {\cal L}^{NN} + {\cal L}^{NY},
   \end{equation}
  where the free ${\cal L}^{\rm free}$ and electromagnetic ${\cal L}^{\rm em}$ parts have the standard form.
%  \begin{eqnarray}
%    {\cal L}^{\rm free} &=& \sum_{B=N, Y}\bar\psi^B ( i\gamma^\mu\partial_\mu - m_B)\psi^B, \\
%    {\cal L}^{\rm em} &=& -\dfrac{1}{4}F^{\mu\nu} F_{\mu\nu} - e\bar\psi\gamma_\mu\frac{1-\tau_3}{2}\psi  A^\mu.
% \end{eqnarray}
% with $\psi^B$ indicating the nucleon ($N$) and hyperon ($Y$) fields,  $m_B$ for the corresponding mass and $F^{\mu\nu}$ for the field tensors of the electromagnetic field $A^\mu$, defined as $F^{\mu\nu}=\partial^\mu A^\nu- \partial^\nu A^\mu$.

 The last two terms ${\cal L}^{NN}$ and ${\cal L}^{NY}$ in Eq. (\ref{Lag1}) describe the effective nucleon-nucleon ($NN$) and nucleon-hyperon $(NY)$ strong interactions and they are parameterized phenomenologically into different forms in different version of RMF approaches. Generally speaking,  these terms can be classified into two types, i.e., the meson-exchange version and the point-coupling version, according to the way how the nucleons and hyperons interact in the hypernuclei.

 \subsubsection{Meson-exchange version}

 In the meson-exchange version of RMF model for hypernuclei, nucleons and hyperons interact effectively via an exchange of mesons ($\phi_m$),
including the scalar $\sigma$, vector $\omega_\mu$, and isovector-vector $\vec\rho_\mu$ mesons, which are employed in order
to simulate different characters of nuclear forces. The electromagnetic interaction between protons is described in terms of photon field $A_\mu$.
The bare charge $e^B$ is adopted for nucleons ($e^n=0, e^p=1$) and $\Lambda$ hyperon ($e^\Lambda=0$). The $NN$ interaction term in Eq.(\ref{Lag1}) has the same form as that for ordinary nuclei (see Chap. 2).

In this model, the Lagrangian density for the $NY$ interaction has the following form
\begin{eqnarray}
\label{FR-NY}
 \mathcal{L}^{NY}
 & = & \bar{\psi}^Y \left(- g_{\sigma Y} \sigma
                          - g_{\omega Y} \gamma^{\mu} \omega_{\mu}
                          + \frac{f_{\omega YY}}{4m_Y}  \sigma^{\mu\nu} \Omega_{\mu\nu}  \right)
             \psi^Y
 \nonumber \\
 &   & +  \mathcal{L}^{\rho Y} + \mathcal{L}^{AY},
 \label{eq:LL}
\end{eqnarray}
where $g_{\sigma Y}$ and $g_{\omega Y}$ are the coupling constants of the hyperon with the scalar and vector meson fields, respectively. The term proportional to $f_{\omega YY}$
with $\sigma^{\mu\nu}=\dfrac{i}{2}[\gamma^\mu, \gamma^\nu]$
represents the tensor coupling between the hyperon and the $\omega$ field.
For $\Lambda$ hyperon ($Y=\Lambda$), since it is
isoscalar and  charge neutral, the last two terms $\mathcal{L}^{\rho Y}$ and $\mathcal{L}^{AY}$
vanish, which describes the couplings of $\Lambda$ with the $\rho$ meson and the photon.  In the following, we will replace the index $Y$ with $\Lambda$ since we mainly
focus on the studies of $\Lambda$ hypernuclei in this Chapter.

Notice that the $NN$ and $NY$ interactions in this model have finite range.
Therefore, the meson-exchange version of RMF model is also referred to as finite-range RMF model.

With the {\em no-sea} and {\em mean-field approximation} at the
Hartree level, the contributions of anti(quasi)-particles and quantum fluctuations of meson fields are neglected and the meson fields are thus
replaced with their expectation values. Following the standard procedure~\cite{Meng06}, one can derive an energy functional corresponding to the Lagrangian density (\ref{Lag1}). Using the Kohn-Sham ansatz, the density of strongly interacting many-body system can be generated from non-interacting many-body system. The equations of motion for the nucleons and hyperons are obtained by minimization of the energy functional with respect the auxiliary single-particle wave function $\psi^B_k$, the bilinear of which gives the densities and currents in the energy functional. The resultant Dirac equation for nucleons in hypernuclei has the same form as that for atomic nuclei and it has been introduced in the previous Chapters. Therefore, here we present the equation only for the $\Lambda$ hyperon, which has the following form
 \begin{eqnarray}
 \left[\mathbf{\alpha}\cdot \mathbf{p}+(m_\Lambda + U_S)
 +  U_V
 -  \frac{f_{\omega \Lambda\Lambda}}{2m_\Lambda} \sigma_{\mu\nu} \partial^\mu \omega^\nu \right]
\psi^\Lambda_k (\bm{r})=\epsilon^\Lambda_k\psi^\Lambda_k (\bm{r}),
 \end{eqnarray}
 with the scalar potential $ U_S =  g_{\sigma \Lambda} \sigma$ and vector potential $\displaystyle  U_V  = g_{\omega \Lambda} \gamma_\mu \omega^\mu$.

 We note that in the finite-range RMF model, the nucleons and hyperon interact
with each other through the exchange of mesons. Therefore, the behavior of nucleons in hypernuclei is modified by the changes in the meson fields due to the presence of hyperons, which provides additional contribution to the sources $S_{\phi_m}$ for the meson field $\phi_m$, fulfilling the Klein-Gordon equation
 \begin{eqnarray}
 \label{MesonEqs}
  \partial_\mu\partial^\mu \phi_m + U^\prime(\phi_m)
  =\pm S_{\phi_m},
 \end{eqnarray}
 where the $(+)$ sign is for vector fields and the $(-)$ sign for
 the scalar field. The source terms $S_{\phi_m}$ in Eq.(\ref{MesonEqs})
 are sums of bilinear products of Dirac spinors
 \begin{eqnarray}
   S_{\phi_m}=
  \left\{\begin{array}{ll}
         \sum\limits_{B,k>0}v^2_k\overline{\psi}^B_k \psi^B_k, & \phi_m=\sigma \\
         \sum\limits_{B,k>0}v^2_k\overline{\psi}^B_k \gamma_\mu\psi^B_k
         - \frac{f_{\omega \Lambda\Lambda}}{2m_\Lambda}j^\Lambda_{T,\mu}, & \phi_m=\omega_\mu\\
         \sum\limits_{B,k>0}v^2_k\overline{\psi}^B_k \gamma_\mu \vec\tau\psi^B_k, & \phi_m=\vec\rho_\mu,
    \end{array}
  \right.
 \end{eqnarray}
 where the sums run over only the positive-energy states ($k>0$) (i.e., no-sea approximation) and
 the occupation probability of the single-particle energy level $k$, i.e., $v^2_k$, is evaluated
 within the BCS or the generalized Bogoliubov transformation method.  The tensor current $j^\Lambda_{T,\mu}$ is defined as,
 \begin{eqnarray}
 j^\Lambda_{T,\mu} = \sum_{\Lambda,k}v^2_k\partial^\nu(\bar\psi^\Lambda_k  \sigma_{\mu\nu}\psi^\Lambda_k).
 \end{eqnarray}

 \subsubsection{Point-coupling version}

 In the point-coupling version of RMF model for hypernuclei, nucleons ($N$) and hyperons ($Y$) interact effectively via contact couplings (zero-range) with different vertices.   

   \begin{figure}[t]
  \centering
  \includegraphics[width=7cm]{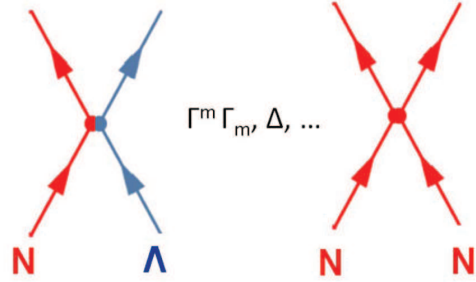}
   \caption{Diagrams for the  nucleon-$\Lambda$ and nucleon-nucleon ($NN$) interactions in the point-coupling RMF model.  }
   \label{diagram}
 \end{figure}
The $NN$ interaction term in Eq.(\ref{Lag1}) has been given in Chap.2. Similarly to the $NN$ interaction, one can construct the point-coupling terms for the $N\Lambda$ interaction,
${\cal L}^{\rm N\Lambda}$. In Ref.~\cite{Tanimura2012Phys.Rev.C14306}, the ${\cal L}^{\rm N\Lambda}$ has been constructed as follows,

\begin{eqnarray}
\begin{aligned}
{\cal L}^{N\! \Lambda}=
{\cal L}_{\rm 4f}^{N\! \Lambda}+{\cal L}_{\rm der}^{N\!\Lambda}
+{\cal L}_{\rm ten}^{N\!\Lambda},
\end{aligned}
%\label{eq:}
\end{eqnarray}
with
 \bsub
\begin{eqnarray}
{\cal L}_{\rm 4f}^{N\!\Lambda}
&=&
-\alpha_S^{(N\!\Lambda)}(\bar{\psi}^{N}\psi^N)(\bar{\psi}^{\Lambda}\psi^{\Lambda})
%\nonumber\\
%&&
-\alpha_V^{(N\!\Lambda)}(\bar{\psi}^{N}\gamma_{\mu}\psi^N)
(\bar{\psi}^{\Lambda}\gamma^{\mu}\psi^{\Lambda}), \\
{\cal L}_{\rm der}^{N\!\Lambda}
&=&-\delta_S^{(N\!\Lambda)}(\partial_{\mu}\bar{\psi}^{N}\psi^N)
(\partial^{\mu}\bar{\psi}^{\Lambda}\psi^{\Lambda}) \nonumber\\
&& -\delta_V^{(N\!\Lambda)}(\partial_{\mu}\bar{\psi}^{N}\gamma_{\nu}\psi^N)
(\partial^{\mu}\bar{\psi}^{\Lambda}\gamma^{\nu}\psi^{\Lambda}),\\
{\cal L}_{\rm ten}^{N\!\Lambda}
&=&
\alpha^{(N\!\Lambda)}_T(\bar{\psi}^{\Lambda}\sigma^{\mu\nu}\psi^{\Lambda})
(\partial_{\nu}\bar{\psi}^{N}\gamma_{\mu}\psi^N).
\end{eqnarray}
\esub
Similar to the finite-range RMF model for hypernuclei, the tensor coupling term ${\cal L}_{\rm ten}^{N\!\Lambda}$, which plays an important role in reproducing the small
spin-orbit splittings in the $\Lambda$ single-particle spectrum~\cite{Noble80,Jennings1990Phys.Lett.B246.325}, has been introduced.

In the point-coupling RMF model, there is no mesonic degree-of-freedom. The nucleons interact with hyperons directly with the
zero-range forces. Therefore, the behavior of nucleons is modified by changes in the mean-field potentials due to the presence of hyperons. This picture is equivalent to that presented in the meson-exchange RMF model in the sense that the meson fields play the same role as
that of mean-field potentials.

The Dirac equation for nucleons in the hypernuclear system reads
\begin{eqnarray}
\label{DiracEq:nucleon}
[\vec{\alpha}\cdot\vec{p}+V_V+V_{TV}\tau_3+V_C+(m_N+V_S+V_{TS}\tau_3)\beta]\psi^N_k=\epsilon_k^N\psi^N_k,
\end{eqnarray}
where the potentials are functionals of various densities and currents. We note that the time-reversal invariance is violated in mean-field approximation for odd-mass systems, for which, one has to take into account the contribution of time-odd fields, namely, the currents generated by the unpaired particles. The two-fold degeneracy of time-reversal single-particle states will be lost.
The study of hypernuclei in the RMF approach with time-odd fields have been carried out only in Ref.~\cite{Sang2013.PhysRevC.88.064304} within a time-odd triaxial RMF approach~\cite{Yao2006Phys.Rev.C24307}, where the time-odd fields separate the single-particle energies of time-reversal partner states by up to 100 keV.
In most calculations,
for the sake of simplicity, an {\em equal-filling approximation}, i.e., the unpaired particle is equally distributed on the time-reversal partner states, is adopted to keep the time-reversal invariance, except in the studies of hypernuclear magnetic moments~\cite{Sang2013.PhysRevC.88.064304}. We will discuss the effects of time-odd fields on hypernuclear magnetic moments in Section~\ref{sec:HyperMM}.

In the following, we adopt the equal-filling approximation if not mentioned explicitly. In this case, the currents vanish and the potentials in (\ref{DiracEq:nucleon}) are simply functionals of densities,
\bsub
\begin{eqnarray}\label{dirac-nucleon}
V_S&=&\delta_S\Delta\rho_S^N+ \alpha_S\rho_S^N
+\beta_S(\rho_S^N)^2+\gamma_S(\rho_S^N)^3
\nonumber\\
&&
+\delta_S^{(N\Lambda)}\Delta\rho_S^\Lambda+\alpha_S^{(N\Lambda)}\rho_S^\Lambda, \\
V_V&=&\delta_V\Delta\rho_V^N+\alpha_V\rho_V^N+
\gamma_V(\rho_V^\Lambda)^3\nonumber\\
&&+\delta_V^{(N\Lambda)}\Delta\rho_V^\Lambda
+\alpha_T^{(N\Lambda)}\rho_T^\Lambda+\alpha_V^{(N\Lambda)}\rho_V^\Lambda, \\
V_{TS}&=&\delta_{TS}\Delta\rho_{TS}^N+\alpha_{TS}\rho_{TS}^N, \\
V_{TV}&=&\delta_{TV}\Delta\rho_{TV}^N+\alpha_{TV}\rho_{TV}^N, \\
V_C&=&eA_0\frac{1-\tau_3}{2},
\end{eqnarray}
\esub
The density $\rho^B_{m=S,TS,V,TV}$ in the different channels and the $\Lambda$-tensor density $\rho^\Lambda_T$ are defined as
\begin{eqnarray}
\rho^B_m=\sum \limits_k\bar\psi^B_k\Gamma_m \psi^B_k, ~~
\rho^\Lambda_T=\nabla\cdot\left(\sum_k\bar\psi^\Lambda_k i\vec{\alpha}\psi^\Lambda_k\right),
\end{eqnarray}
where $\Gamma_m$ is $1, \tau_3, \gamma_0, \gamma_0\tau_3$, respectively.

The Dirac equation for $\Lambda$ hyperon reads
 \begin{eqnarray}
 \label{DiracEq:hyperon}
[\vec{\alpha}\cdot\vec{p}+U_V+ U_T+(U_S+m_\Lambda)\beta]\psi^\Lambda_k
=\epsilon_k^\Lambda\psi^\Lambda_k,
\end{eqnarray}
with the scalar, vector and tensor potentials given by
\bsub\begin{eqnarray}\label{dirac-Lambda}
U_S&=&\delta_S^{(N\Lambda)}\Delta\rho_S^N+\alpha_S^{(N\Lambda)}\rho_S^N,\\
U_V&=&\delta_V^{(N\Lambda)}\Delta\rho_V^N+\alpha_V^{(N\Lambda)}\rho_V^N,\\
U_T&=&-i\alpha_T^{(N\Lambda)}\beta\vec{\alpha}\cdot(\vec{\nabla}\rho_V^N).
\end{eqnarray}
\esub
 The Dirac equations for nucleons (\ref{DiracEq:nucleon}) and hyperons (\ref{DiracEq:hyperon}) are solved iteratively until a convergence is reached with a desired
precision.

The RMF energy $E_{\rm RMF}$ for a single-$\Lambda$ hypernucleus ($^{A+1}_{~~~\Lambda}Z$ composed of $A$ nucleons and one $\Lambda$) is determined as
\begin{eqnarray}
\label{energy:hypernucleus}
E_{\rm RMF}&=& \sum_{k=1} v^2_k\epsilon_k+\epsilon_{\Lambda}-Am_N-m_{\Lambda} \nonumber\\
&&-\int d^3\br\ \biggl[
\frac{1}{2}\sum_{m}(\alpha_m(\rho^N_m)^2+\delta_m\rho^N_m\Delta\rho^N_m)
+\frac{1}{2}eA^0\rho^{(p)}_V \biggl.\nonumber\\
&&\biggl.
+\frac{2}{3}\beta_S(\rho^N_S)^3+\frac{3}{4}\gamma_S(\rho^N_S)^4+\frac{3}{4}\gamma_V(\rho^N_V)^4
\biggr]\nonumber\\
&&
-\int d^3\br  \biggl[\sum\limits_{m=S,V}(\alpha_m^{(N\!\Lambda)}\rho^N_m\rho_m^{\Lambda}
+ \delta_m^{(N\!\Lambda)}\rho^N_m\Delta\rho^{\Lambda}_m)
+\alpha^{(N\!\Lambda)}_T\rho^N_V\rho_T^{\Lambda}\biggr], \nonumber \\
\end{eqnarray}
where the summation over $m$ in the second line runs over all 
the channels, $S, V, TS$, and $TV$. Alternatively, the energy of single-$\Lambda$ hypernucleus (\ref{energy:hypernucleus}) can also be evaluated as a sum of kinetic energy of nucleons $T_N$ and hyperons $T_\Lambda$, $NN$ interaction energy $E_{NN}$, and $N\Lambda$ interaction energy $E_{N\Lambda}$ terms 
\begin{eqnarray}
E_{\rm RMF}&=& T_N + T_\Lambda + E_{NN} + E_{N\Lambda} \nonumber \\
&=&\int d^3\br\biggl[
\sum\limits_{k=1}^{N}v^2_k\psi^\dagger_k(\vec{\alpha}\cdot\vec{p}+m_N\beta-m_N)\psi_k+
\psi^\dagger_\Lambda(\vec{\alpha}\cdot\vec{p}+m_\Lambda\beta-m_\Lambda)\psi_\Lambda\biggl]\nonumber \\
&& +\int d^3\br \biggl[\frac{1}{2}\sum\limits_m(\alpha_m(\rho_m^N)^2+ \delta_m\rho^N_m\Delta\rho_m^N)+\frac{1}{2}eA^0\rho^{(p)}_V  \biggl.\nonumber\\
&&\biggl.+\frac{1}{3}\beta_S(\rho_S^N)^3+\frac{1}{4}\gamma_S(\rho_S^N)^4
+\frac{1}{4}\gamma_V(\rho_V^N)^4\biggl] \nonumber\\
&&+\int d^3\br\biggl[\sum\limits_{m=S,V}(\alpha_m^{(N\Lambda)}\rho_m^N\rho_m^\Lambda+\delta_m^{(N\Lambda)}\rho_m^N\Delta\rho_m^\Lambda)+
\alpha_T^{(N\Lambda)}\rho^N_V\rho_T^\Lambda\biggl].
\end{eqnarray}

The center-of-mass correction (CM) energy $E_{\rm CM}$ is calculated by taking the expectation value of the kinetic energy for the center of mass motion with respect to the many-body ground state wave function. For a single-$\Lambda$ hypernucleus, it is given by
\begin{equation}
E_{\rm CM}=\frac{\langle \vec{P}_{\rm CM}^2 \rangle}{2[Am_N+m_{\Lambda}]},
\label{cmenergy}
\end{equation}
 where $\vec{P}_{\rm CM}$ is the total momentum of a hypernucleus with $A$ nucleons and one $\Lambda$ hyperon.

 For $\Lambda$ hypernuclei with open-shell nuclear core, the
pairing correlation among
nucleons needs to be taken into account. In this case, the pairing energy $E_{\rm pair}$ related to pairing density should also be added to the total energy. Therefore, the total binding energy $B (^{A+1}_\Lambda Z)$ of a single-$\Lambda$ hypernucleus is finally given by
 \beq
 B (^{A+1}_\Lambda Z) = - (E_{\rm RMF}-E_{\rm CM} + E_{\rm pair}).
 \eeq

\section{Ground state properties of hypernuclei}
\subsection{Systematics of $\Lambda$ binding energy}

Let us now discuss applications of the relativistic approaches to the ground
state properties of hypernuclei. We first discuss the systemtatics of binding energy
of $\Lambda$ particle in single-$\Lambda$ hypernuclei.

The $\Lambda$ binding energy is defined as
\begin{eqnarray}
B_\Lambda&=&M(^AZ)c^2+m_\Lambda c^2-M(^{A+1}_{~~~\Lambda} Z)c^2, \\
&=&B(^{A+1}_{~~~\Lambda} Z)-B(^AZ),
\end{eqnarray}
where $M(^{A+1}_{~~~\Lambda} Z)c^2$ and $M(^AZ)c^2$ are the mass of a hypernucleus and that of the
corresponding core nucleus, respectively, while $m_\Lambda c^2$ is the mass of a $\Lambda$ particle (see Table \ref{table1}).
$B(^{A+1}_{~~~\Lambda} Z)$ and $B(^AZ)$ are the binding energy of the hypernucleus and the
core nucleus, respectively.
The binding energies for light hypernuclei with $\Lambda$
particle in the 1$s$ state (that is, the lowest single-particle state) were determined by early emulsion experiments ~\cite{Davis92}.
The $\Lambda$ binding energies for higher $l$ single-particle states, from light to medium-heavy and heavy hypernuclei, such as $^{28}_{~\Lambda}$Si, $^{89}_{~\Lambda}$Y,
$^{139}_{~~\Lambda}$La, and $^{209}_{~~\Lambda}$Pb, have been obtained with the ($\pi^+,K^+$)
reaction spectroscopy ~\cite{HT06}.
More recently, the ($e,e'K^+$) reaction spectroscopy has also been
developed~\cite{Nue13,Gogami14}, which provides
a higher energy resolution than the ($\pi^+,K^+$) reaction spectroscopy.
One of the most important achievements with these reaction spectroscopies is that
single-particle $\Lambda$ states have been clearly identified in hypernuclei from
$s$ ($l$=0) to $h$ ($l$=5) states ~\cite{HT06}.

\begin{figure}[t]
%\centering
\begin{center}
\includegraphics[width=7cm]{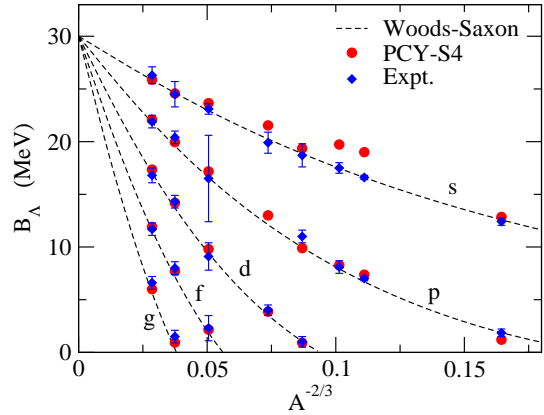}
\caption{The mass number dependence of $\Lambda$ binding energies plotted as a function of
$A^{-2/3}$. The dashed lines are obtained with a Woods-Saxon potential with
a depth parameter of 30 MeV, the diffuseness parameter of 0.6 fm, and the radius parameter of 1.1 fm.
The filled circles show the results of spherical mean-field calculations with the relativistic
point coupling model with the PCY-S4 parameter set ~\cite{Tanimura2012Phys.Rev.C14306}.
The filled diamonds show the experimental data taken from Ref. ~\cite{HT06} (see also Table I in
Ref. ~\cite{UB99}).
}
\label{fig-spe}
\end{center}
\end{figure}

The mass number dependence of $\Lambda$ binding energies have been well fitted with a simple Woods-Saxon potential for a $\Lambda$ particle~\cite{MDG88,MBWZ88}.
See the dashed lines in Fig. \ref{fig-spe} for a fit with a Woods-Saxon potential in which the depth, the diffuseness, and the radius parameters are $V_0=-30$ MeV, $a$=0.6 fm, and
$R=1.1\cdot A^{1/3}$ fm, respectively ~\cite{HT06,MBWZ88}.
If one approximates a Woods-Saxon potential with
an infinite square well potential with the depth of $-V_0$ and the range of $R$,
the energy of the 1$s$ state reads
\begin{equation}
\epsilon_{1s}=-V_0+\frac{\hbar^2}{2m_\Lambda}\cdot\frac{\pi^2}{R^2}.
\label{eq-spe}
\end{equation}
If one assumes that the depth parameter $V_0$ is independent of the mass number of hypernucleus, $A$, and the range parameter $R$ is proportional to $A^{1/3}$, this equation
implies that the $\Lambda$ single-particle energies scale as $A^{-2/3}$. This has indeed been
observed experimentally, at least for medium-heavy and heavy hypernuclei, as can be seen in
Fig. \ref{fig-spe}. The $\Lambda$ binding energies are therefore often plotted as a
function of $A^{-2/3}$.
Moreover, Eq. (\ref{eq-spe}) suggests that the depth parameter $V_0$ can be obtained by
extrapolating the single-particle energies to $A=\infty$.
The empirical value of the depth parameter is
about $V_0=30$ MeV (see Fig. \ref{fig-spe}).
This value is about 2/3 of the depth parameter for nucleons, indicating
that the $\Lambda N$ interaction is weaker than the $NN$ interaction.
A similar value of the depth parameter was obtained earlier in Refs.
~\cite{Boussy77,Boussy79,Boussy80}.

A microscopic understanding of the $\Lambda$ binding energies can be achieved with
the self-consistent mean-field approaches. As a matter of fact, the experimental $\Lambda$ binding
energies have been used as important empirical inputs
in determining the value of
parameters of an effective $N\Lambda$ interaction for mean-field calculations.
This has been done both for the non-relativistic Skyrme-Hartree-Fock apprach ~\cite{YBZ88}
and for the relativistic mean-field approach ~\cite{MZ89,ST94,RSM90,Song2010PK1-Y1,Tanimura2012Phys.Rev.C14306}.
As an example, Fig. \ref{fig-spe} shows the results of spherical mean-field
calculations with the PCY-S4 parameter set for the
relativistic point-coupling Lagrangian ~\cite{Tanimura2012Phys.Rev.C14306}. One can see that
the overall fit is good except for a few deformed hypernuclei.

\subsection{Spin-orbit splittings}

We next discuss the spin-orbit splittings of hypernuclei.
In ordinary nuclei, large spin-orbit splittings are one of the most
important quantities in order to understand the shell structures of atomic nuclei.
In contrast, in hypernuclei, the ($K^-,\pi^-$) reaction spectroscopies
have revealed already by the late 1970's that the spin-orbit splittings are much
smaller~\cite{BFK78}. More recently, the $\gamma$-ray spectroscopy has been carried out
for the $^{13}_{~\Lambda}$C hypernucleus, which has provided a more quantitative information on
the spin-orbit splittings~~\cite{Ajimura2001}. In the experiment, $\gamma$-rays from
the excited $1/2^-$ and $3/2^-$ states to the ground state were measured following
the $^{13}$C($K^-$,$\pi^-$)$^{13}_{~\Lambda}$C reaction~~\cite{Ajimura2001}.
From this experiment, the spin-orbit splitting between the 1$p_{1/2}$ and the 1$p_{3/2}$ hyperon states
in $^{13}_{~\Lambda}$C was determined to be $152\pm54 \pm36$~keV,
which is smaller than the spin-orbit splitting in ordinary nuclei by a factor of 20-30~\cite{HT06,Ajimura2001}.

As we have mentioned in the Introduction, the early applications the relativistic approaches
focused on discussions on the spin-orbit splittings.
Brockmann and Weise first argued that the particular structure of the 2$\pi$ and 3$\pi$ exchange interactions
lead to a reduction of
$N\Lambda$ coupling constants from the corresponding
$NN$ coupling constants by a factor of about 3, considerably quenching the
spin-orbit splittings ~\cite{BW77} (notice that soon after the publication of Brockmann and Weise, Pirner pointed out that the scaling factor is about 2/3 rather
than 1/3 if the quark model is adopted~\cite{Pirner79}). See also Ref. ~\cite{BB81} for a similar argument.
Brockmann and Weise also investigated
the effect of Fock exchange terms on the spin-orbit splittings~\cite{BW81}. They have shown that
the K and K$^*$ meson exchange contributions almost cancel with each other, and have concluded that
the essential features of the spin-orbit splittings
are well represented already in the Hartree approximation.
Subsequently, Noble pointed out the importance of the tensor coupling (see Eq. (\ref{FR-NY})),
\begin{equation}
{\cal L}_{\rm tensor}=\frac{f_\omega}{2m_\Lambda}\bar{\psi}_\Lambda\,\sigma^{\mu\nu}\psi_\Lambda\,\partial_\nu\omega_\mu,
\end{equation}
in the spin-orbit splittings~\cite{Noble80}. That is, Noble showed that the contribution of the
tensor coupling term largely cancels the Thomas precession term of the spin-orbit potential, resulting in
the small spin-orbit splittings in $\Lambda$ hypernuclei. The tensor coupling has further been investigated {\it e.g.,} in Refs.~\cite{Boussy81,Boussy82,ZQ85,ST94,Tanimura2012Phys.Rev.C14306},
and its role in the spin-orbit splittings has by now been well established.

\subsection{Hypernuclear magnetic moments}\label{sec:HyperMM}

 The magnetic moments of hypernuclei are important physics observables, since they are sensitive to the spin and angular momentum structure of hypernuclei as well as the
spin-dependent hyperon-nucleon interactions. In particular, these quantities provide direct information on the properties of hadrons in the nuclear medium.

 The first study of hypernuclear magnetic moments was performed for light p-shell hypernuclei within a three-body cluster model~\cite{Motoba85}. Thereafter, magnetic moments of light hypernuclei were theoretically studied within the shell model, and predicted to be close to the Schmidt lines~\cite{Tanaka1989}, which are obtained with
extreme single-particle model neglecting the core polarization due to a particle or a hole added.
However, the magnetic moments of hypernuclei may deviate
from the Schmidt values, especially if the
meson ($K$ and $\pi$) exchange currents are considered~\cite{Saito1997}.

\begin{figure}[h!]
  \centering
  \includegraphics[width=8cm]{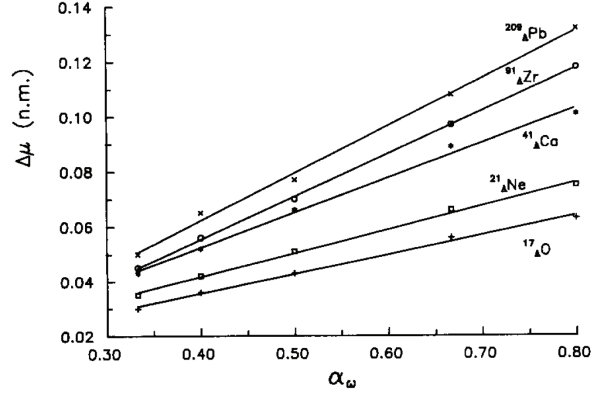}
   \caption{The dependence of the deviation of a magnetic moment from the Schmidt value,
$\Delta\mu$, on the ratio of the vector meson coupling constants,
$\alpha_\omega(\equiv g_{\omega\Lambda}/g_{\omega N})$. Magnetic moments (with $\Lambda$ in the $s_{l/2}$ state) are calculated within the RMF model without the non-linear couplings of mesons  for the $NN$ part and several pairs of ratios of coupling constants $(\alpha_\sigma, \alpha_\omega)$ for the $N\Lambda$ interaction. The figure is taken from Ref.~\cite{Mares1990}. }
   \label{magneticmoment}
 \end{figure}

 In the RMF description of magnetic moments in ordinary nuclei near double-closed shell, the contribution of the polarized current from core nucleons is compensated by the enhanced Dirac current from the valence charged particle due to the reduction of the nucleon mass~\cite{Yao2006Phys.Rev.C24307}. However, this cancelation is not expected in $\Lambda$-hypernuclei due to the charge neutrality of the $\Lambda$ hyperon. Therefore, the polarized proton current induced by the valence hyperon causes the total magnetic moment to deviate from the Schmidt value. Fig.~\ref{magneticmoment} shows the dependence of the deviation, $\Delta\mu$,
on the ratio of the vector meson coupling constants,
$\alpha_\omega(\equiv g_{\omega\Lambda}/g_{\omega N})$. It is seen that the deviations are proportional to $\alpha_\omega$.
These deviations were suggested as an indicator of relativistic effects in nuclei~\cite{Cohen1987PhysRevC.35.2231,Mares1990}. However, it has been eventually realized that the tensor coupling of vector field to the $\Lambda$ hyperon could significantly change the hypernuclear magnetic moments by renormalizing the electromagnetic current vertex in nuclear medium, and bring
 the magnetic moment of $\Lambda$-hypernucleus with $\ell_\Lambda=0$ close to the Schmidt value again, although this is not the case for $ \ell_\Lambda\ne0$
~\cite{Gattone1991PhysRevC.44.548,Yao2008}.

%-------------------------------------------
  \begin{table}[h!]
   \centering
   \tabcolsep=9pt
    \tbl{Magnetic moments of $\Lambda$ hypernuclei $^{17}_{~\Lambda}$O and
$^{41}_{~\Lambda}$Ca.
All results are taken from Ref.~\cite{Yao2008}.  }
   {\begin{tabular}{ cccc|ccc}
  \hline\hline
       &    \multicolumn{3}{c|}{$^{17}_\Lambda$O}& \multicolumn{3}{c}{$^{41}_\Lambda$Ca }\\
   \cline{2-7}
       $\mu$ (n.m.)         &  $1s_{1/2}$ &  $1p_{3/2}$ & $1p_{1/2}$&  $1s_{1/2}$ &  $1p_{3/2}$ &  $1p_{1/2}$     \\
   \hline

     Schmidt                & -0.613      & -0.613      &  0.204    & -0.613      & -0.613      &   0.204 \\
      RMF                  & -0.660      & -0.662      &  0.170    & -0.682      &  -0.718     &   0.153   \\
      RMF+Tensor             & -0.611      & -0.643      &  0.186    & -0.611      &  -0.667     &   0.170  \\
 \hline\hline
  \end{tabular}}
  \label{tab1}
 \end{table}

 It is shown in Table~\ref{tab1} that there is an evident difference between the magnetic moments by the RMF and the Schmidt formula, which mainly comes from the polarized Dirac magnetic moments.  After taking into account the tensor
 effect on the current, the difference almost disappears for the $\Lambda$ hyperon in the $1s$ state.
For the $1p$ states,
this difference is greatly suppressed, but there is still a $5\%-25\%$ difference for a $\Lambda$ hyperon in the $1p$ state.

\begin{figure}[]
 \centering
  \includegraphics[width=8cm]{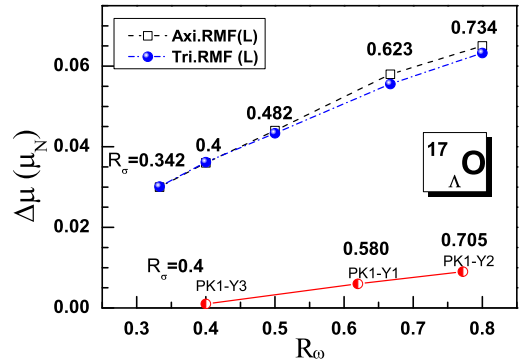}
   \caption{Same as Fig.~\ref{magneticmoment}, but only for $^{17}_{~\Lambda}$O obtained with  the time-odd axial (labeled with open squares, taken from Ref.~\cite{Mares1990}) and with the  time-odd triaxial RMF (labeled with bullets) calculations. The results from the time-odd triaxial RMF calculation with the $\omega\Lambda\Lambda$ tensor coupling are also plotted with half-filled circles. Here, $R_\sigma\equiv g_{\sigma \Lambda}/g_{\sigma N}\omega$ and $R_\omega\equiv g_{\omega \Lambda}/g_{\omega N}$. This figure is taken from Ref.~\cite{Sang2013.PhysRevC.88.064304}.}
   \label{magnetic_moment2}
 \end{figure}

 We note that the magnetic moments of $\Lambda$ hypernuclei from light to heavy mass are systematically investigated recently based on the perturbation consideration of core polarization effect within the spherical RMF model~\cite{Sang2013epja} as well as the self-consistent solution of the time-odd triaxial RMF model~\cite{Sang2013.PhysRevC.88.064304}. Figure~\ref{magnetic_moment2} displays the dependence of $\Delta\mu$ on the ratio of meson-hyperon coupling strengths $R_\sigma\equiv g_{\sigma \Lambda}/g_{\sigma N}\omega$ and $R_\omega\equiv g_{\omega \Lambda}/g_{\omega N}\omega$ for $^{17}_{~\Lambda}$O hypernucleus. As expected, the time-odd triaxial RMF calculation gives similar results as that of time-odd axial RMF calculation for $^{17}_{~\Lambda}$O if the same effective interaction is used. In both calculations, the vector meson tensor coupling is not taken into account. Figure~\ref{magnetic_moment2} also shows the results by the time-odd triaxial RMF calculation using the parameter set PK1-Y1, PK1-Y2 and PK1-Y3~\cite{Wang2013ctp}.
 After taking into account the tensor coupling term, the deviation still increases with the vector coupling strength $g_{\omega\Lambda}$, but with a smaller slope. The self-consistent deformed RMF calculations with time-odd components have confirmed
 the conclusion from the analytical expression for spherical nuclei.
Moreover, Fig.~\ref{magnetic_moment2} indicates that the deviation $\Delta\mu$ also depends on the scalar coupling strength $g_{\sigma\Lambda}$ through the effective scalar mass $M^*$~\cite{Cohen1987PhysRevC.35.2231}.

 Compared with the magnetic moments, the spin-flip $M1$ transitions between hypernuclear doublet states are relatively easier to be measured. A few data have already been accumulated~\cite{Tamura2000PhysRevLett.84.5963,Hashimoto2006}. Therefore,
it is an important future work to extend the RMF approach to study the $M1$ transition strengths in $\Lambda$ hypernuclei.  

\subsection{Stabilization of neutron-rich nuclei on the drip line}

\begin{figure}[tb]
%\centering
\begin{center}
\includegraphics[angle=270,width=11cm]{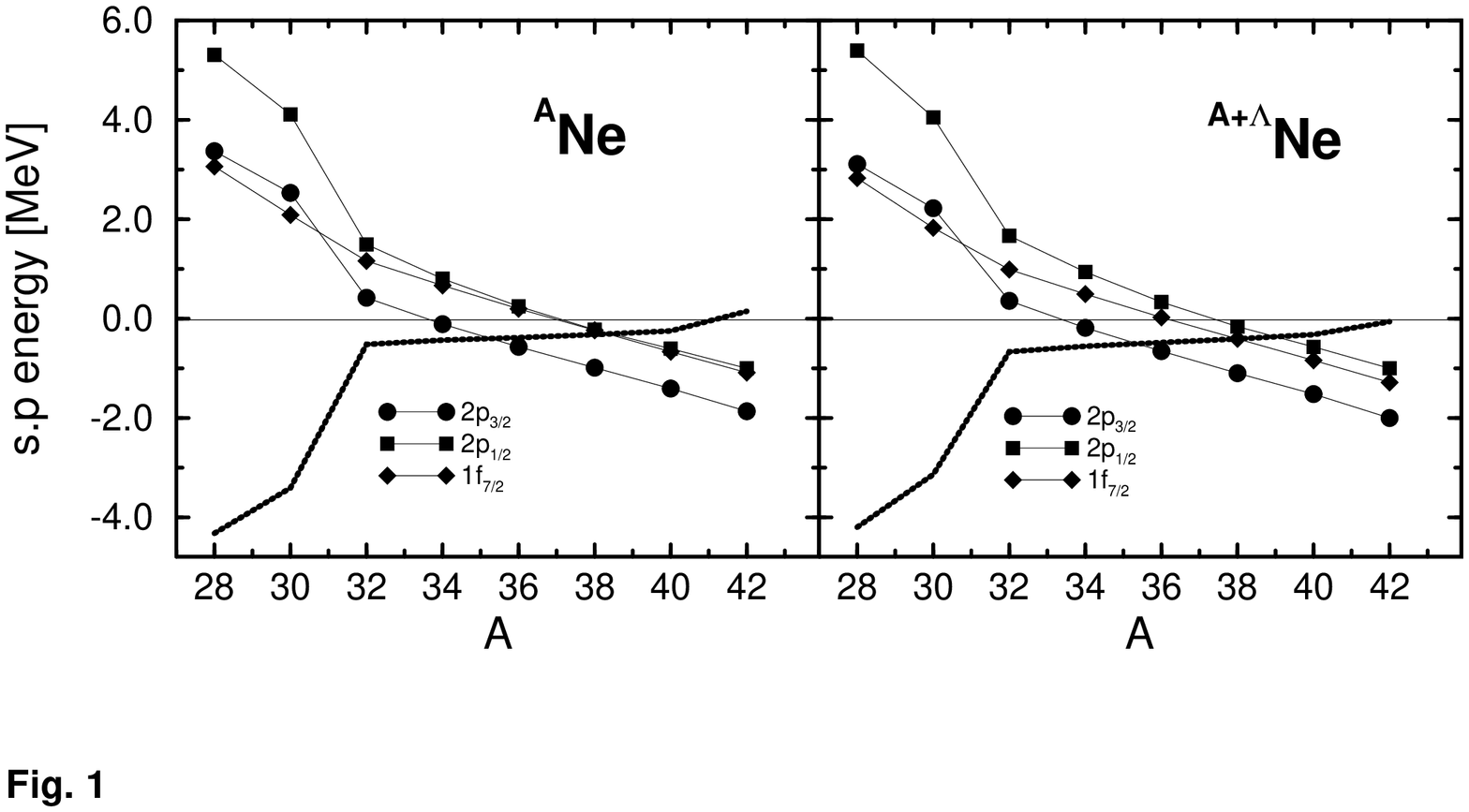}
\caption{Single-particle energies for neutron levels in Ne isotopes (the left panel)
and Ne hypernuclei (the right panel) obtained with the relativistic Hartree-Bogoliubov
method. The thick solid lines denote the Fermi energy. Taken from Ref. ~\cite{VPLR98}.}
\label{fig-vretenar}
\end{center}
\end{figure}

Physics of unstable nuclei is one of the most important topics in the
current nuclear physics ~\cite{TSK13,HTS13,KY12,IMKT10}.
Many interesting phenomena have been found in neutron-rich nuclei, such as
a halo and skin structure, soft dipole excitations, and changes of magic numbers, all
of which originate from the weakly bound character of neutron-rich nuclei.
In particular, the halo structure is one of the most important phenomena in neutron-rich
nuclei, in which the density distribution is largely extended up to large distances.

An interesting question is how the halo structure of neutron-rich nuclei
is influenced by adding a $\Lambda$ particle. This has been studied in Ref. ~\cite{HKMYY96}
using a three-body model, in which the authors have shown that the $\Lambda$ particle
induces an extra binding and as a consequence the halo structure disappears.
In terms of mean-field theory, the extra $\Lambda$ particle leads to a modification of
single-particle potentials for nucleons. Vretenar {\it et al.} have used the relativistic
Hartree-Bogoliubov method and demonstrated that the change in the spin-orbit potential plays an
important role in stabilizing neutron-rich nuclei~\cite{VPLR98}.
This is one of the most successful applications
of the relativistic approach, since the conclusion is difficult to reach with the non-relativistic
approaches, in which the spin-orbit potential is treated phenomenologically.
Vretenar {\it et al.} have also shown that the stabilization of neutron-rich nuclei
leads to an extention of the neutron-drip line ~\cite{VPLR98} (see Fig. \ref{fig-vretenar}).
A similar study has been carried out also by L\"u {\it et al.} ~\cite{LMZZ03}
(see also Ref.~\cite{ZPSV08} for a study with a non-relativistic approach).

\subsection{Deformation of hypernuclei}

It is well known that many open-shell nuclei are deformed in the
ground state. A clear experimental evidence for nuclear deformation is provided by
a rotational spectrum, which scales as $E_I\propto I(I+1)$ as a function of
the angular momentum $I$, as well as by strongly enhanced electric quadrupole transition probabilities.
Theoretically, a standard means to discuss nuclear deformation is a self-consistent mean-field theory.
In working with the intrinsic (body-fixed) frame, in which the rotational symmetry is broken in the mean-field potential,
the mean-field theory provides an intuitive and transparent view of the nuclear deformation.

The mean-field approach is a suitable tool also to investigate how the nuclear deformation is affected by a
$\Lambda$ particle, because the optimum density distribution is obtained automatically by
minimizing the total energy. \v Zofka was the first person who applied the self-consistent method to deformed
hypernuclei ~\cite{Z80}. He used non-relativistic Gaussian interactions for nucleon-nucleon ($NN$) and
nucleon-Lambda ($N\Lambda$) interactions, and showed that a $\Lambda$ particle changes
the quadrupole moment at most by 5 \% in the $sd$-shell region. This result has been confirmed with more recent Skyrme-Hartree-Fock (SHF) calculations for axially deformed hypernuclei~\cite{ZSSWZ07}.

\begin{figure}[bt]
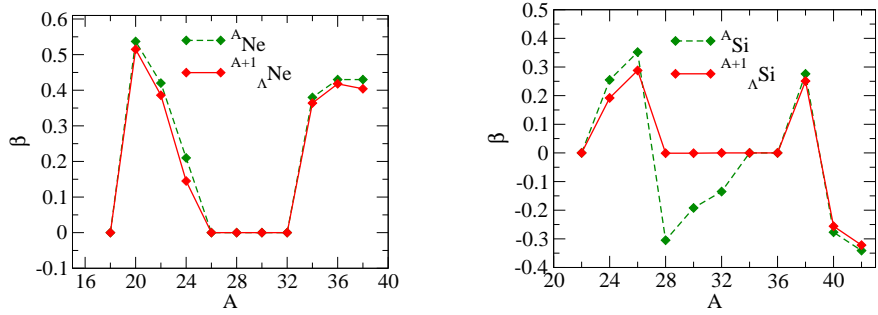

\sidebyside
{\includegraphics[width=.45\textwidth,clip]{fig7-6-1.eps}}
{\includegraphics[width=.45\textwidth,clip]{fig7-6-2.eps}}
\caption{
Quadrupole deformation parameter $\beta$
for Ne (left panel) and Si (right panel)
isotopes obtained with the RMF method with the NL3 parameter set.
The dashed lines show the deformation parameter for the core nucleus, while
the solid lines for the corresponding hypernucleus.
}
\label{def-nesi}
\end{figure}

An application of the relativistic mean field approach to deformed hypernuclei
was first carried out in Ref. ~\cite{WH08}. In these calculations, axial symmetry for the density distribution
was assumed, and a $\Lambda$ particle was put in the lowest single-particle orbit.
The NL3 parameter set of RMF ~\cite{nl3} was employed. Figure \ref{def-nesi} shows the quadrupole
deformation parameter $\beta$ for the ground state of Ne and Si isotopes obtained with these calculations, as well as that for the ground state of corresponding hypernuclei.
The
deformation parameter is defined as
\begin{equation}
\beta\equiv\dfrac{4\pi}{3AR^2} Q_{20},
\end{equation}
with the expectation value of mass quadrupole moments,
\beq
 Q_{20}= \sqrt{\dfrac{5}{16\pi}} \int d^3\br  (2 z^2 - x^2 - y^2) \rho_{\rm tot}(\br),
\eeq
where the total density $\rho_{\rm tot}(\br)$ is contributed from neutrons and protons for ordinary nuclei, and additionally from hyperons for hypernuclei.

One can see that the change in the deformation parameter
for most of the nuclei shown in the figure is small, as in
the non-relativistic self-consistent
calculations ~\cite{Z80,ZSSWZ07}.
However, in the relativistic approach, certain nuclei, that is,
$^{28,30,32}$Si, show a drastic change of the deformation parameter,
from oblate deformation
to spherical when a $\Lambda$ particle is added.
It was shown that a similar disappearance of
nuclear deformation due to a $\Lambda$ particle
also takes place in $^{12}$C, both with RMF~\cite{WH08} and
anti-symmetrized molecular dynamics (AMD)~\cite{IKDO11}.

The disppearance of nuclear deformation was not observed in the
non-relativistic Skyrme-Hartree-Fock (SHF) calculations ~\cite{ZSSWZ07}.
In Ref. ~\cite{SWHS10}, it has been shown that the difference comes about
because the RMF yields a somewhat stronger polarization
effect of the $\Lambda$ hyperon than that of the SHF approach.

\begin{figure}[bt]
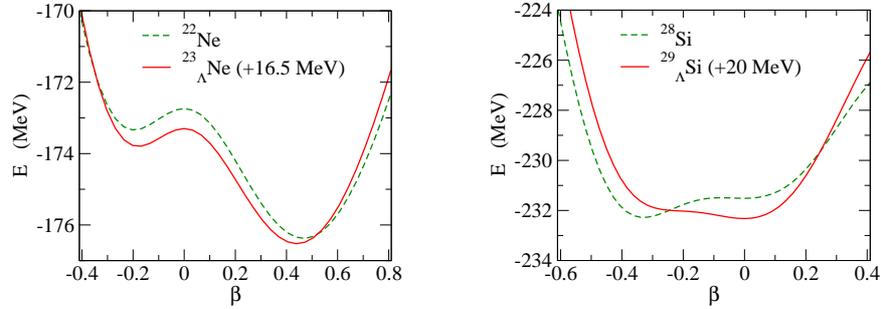

\sidebyside
{\includegraphics[width=.45\textwidth,clip]{fig7-7-1.eps}}
{\includegraphics[width=.45\textwidth,clip]{fig7-7-2.eps}}
\caption{(Left panel)
The potential energy surface for $^{22}$Ne (the dashed line) and
$^{23}_{~\Lambda}$Ne (the solid line) nuclei obtained with the RMF method.
The energy surface for $^{23}_{~\Lambda}$Ne is shifted by a constant amount
as indicated in the figure. (Right panel) The same as the left panel, but
for $^{28}$Si and $^{29}_{~\Lambda}$Si nuclei. }
\label{def-nesi2}
\end{figure}

Figure \ref{def-nesi2} shows
the potential energy surfaces for the
$^{23}_{~\Lambda}$Ne and $^{29}_{~\Lambda}$Si nuclei obtained with the RMF method.
The energy surfaces for the corresponding core nuclei are also shown.
In contrast to the $^{22}$Ne nucleus, which has a deep prolate minimum
in the energy surface,
the energy surface for the $^{28}$Si nucleus shows a
relatively shallow oblate minimum,
with a shoulder at the spherical configuration.
Notice that the $\Lambda$ particle gains the largest binding energy at the
spherical configuration as the overlap with the core nucleus is maximum.
The energy difference between the oblate and the spherical configurations in $^{28}$Si
is as small as 0.754 MeV, and could be easily inverted when a $\Lambda$ particle is
added. This explains why the disappearance of deformation occurs in $^{28}$Si
but not in $^{22}$Ne.

Those calculations have been extended in Ref. ~\cite{LZZ11} by including the triaxial degree
of freedom (see Ref. ~\cite{WHK11} for a similar 3D calculation with the non-relativistic
approach). The main conclusion remained the same as in Ref. ~\cite{WH08}, although the results
are slightly different in details. A general conclusion is that the potential energy surface
tends to be softer both in the $\beta$ (the axial symmetric deformation) and
the $\gamma$ (the triaxial deformation) directions when a $\Lambda$ particle is embedded in a nucleus.
This would indicate that the shape fluctuation effect, that is, the effect beyond the
mean-field approaximation, is more important in hypernuclei as compared to ordinary nuclei.
We will discuss this in the next section.

More recently, Lu {\it et al.} applied the RMF method to superdeformation of hypernuclei ~\cite{LHSZ14}.
They pointed out that there are a few nuclei which show a bubble-like density distribution for
the superdeformed configuration and that the $\Lambda$ binding energy becomes larger at the
superdefomed state compared to that at a normal deformation in that situation.
This implies that the density distribution of
the superdeformed state of ordinary nuclei might be able to be studied by measuring the $\Lambda$ binding energy.

\section{Collective excitations}

\subsection{Impurity effect of Lambda hyperon on nuclear collective excitations}

Let us next consider the collective excitations of hypernuclei.
To this end, it is important to take into account the effect beyond the
mean-field approximation. We first discuss
the impurity effect of $\Lambda$ hyperon on nuclear collective excitation.
In Ref. ~\cite{YLHWZM11},
the collective excitations of nuclear core in $\Lambda$ hypernuclei including rotation and vibration of quadrupole deformed shapes have been described
with a five-dimensional collective Bohr Hamiltonian (5DCH), in which the moments of inertia and mass parameters are functions of deformation parameters ($\beta,\gamma$) determined by self-consistent mean-field calculations. The presence of a $\Lambda$ particle in the atomic nucleus modifies
the mean-field potentials for nucleons and thus changes the energies and wave functions of single-particle states. These changes will enter into the mass parameters in the 5DCH, which are usually calculated with the cranking approximation.
Subsequently, a similar 5DCH study has also been carried out based on the solutions of a triaxially deformed RMF approach for $\Lambda$ hypernuclei~\cite{Xue2014}.

 \begin{figure}[]
 \centering
  \includegraphics[width=11cm]{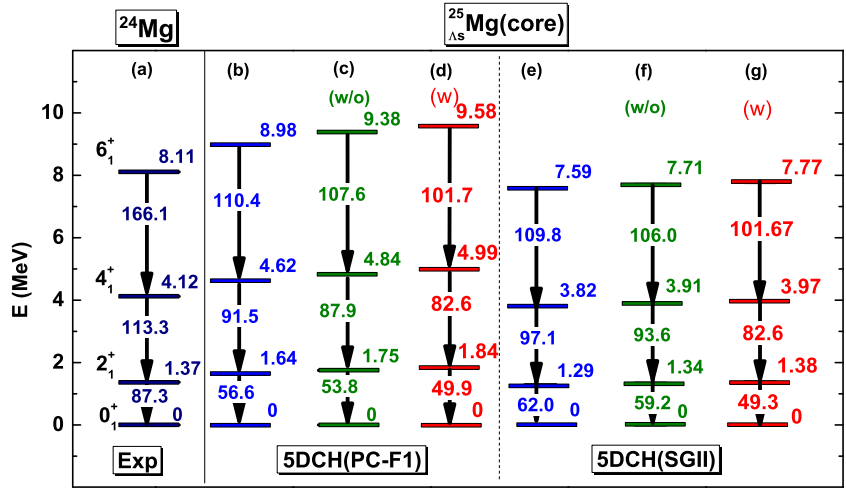}
   \caption{The low-spin spectra of the ground-state rotational band for $^{24}$Mg (a,b,e) and for the nuclear core of $^{25}_\Lambda$Mg (c, d, f, g). The theoretical results are obtained by the 5DCH calculation with the parameters determined by the triaxial RMF (b, c, d) and by the  Skyrme-Hartree-Fock+BCS (e, f, g) calculations. The difference between (c) and (d), and between (f) and (g) is wether the $\Lambda N$ interaction term is included (w) or not (w/o) in the potential energy for the nuclear core part. The $B(E2)$ values are in units of e$^2$ fm$^4$. }
 \label{Mg24_spectrum1}
 \end{figure}

 Figure~\ref{Mg24_spectrum1} displays the low-spin spectra of the ground-state band for $^{24}$Mg and for the
nuclear core of $^{25}_\Lambda$Mg calculated by the 5DCH based on the
relativistic point-coupling EDF with PC-F1 parameter set and the
non-relativistic EDF with the SGII parameter set. It is noted that the $\Lambda$ stretches the spectra of the ground-state band. Comparing
between the
columns (b) and (d), one finds that the $\Lambda$ increases the excitation energy of $2^+_1$ state, $E_x(2^+_1)$,  by $\sim 12\%$ and reduces the $B(E2: 2^+_1 \rightarrow 0^+_1)$ value by $\sim 12\%$. The reduction effect in the relativistic calculation is much larger than that found in the non-relativistic calculation~\cite{YLHWZM11}, as illustrated in the comparison of columns (e) and (g). It is shown that the $\Lambda$ increases $E_x(2^+_1)$ by $\sim 7\%$ and reduces the $B(E2: 2^+_1 \rightarrow 0^+_1)$ by $\sim 9\%$. Considering the fact
that the magnitude of impurity effect of $\Lambda$ hyperon on nuclear collective properties is not much dependent on the underlying EDF used to determine the parameters of the 5DCH~\cite{Mei2012}, one may ascribe this significant difference in the results by the 5DCH(PC-F1) and 5DCH(SGII) to the relativistic effect, which enhances the impurity effect in atomic nuclei. Moreover, the comparison of columns (c) and (d), as well as (f) and (g) demonstrates the important influence of $\Lambda N$ interaction on nuclear collective excitations.

\subsection{Microscopic description of low-energy spectra in Lambda hypernuclei}

We next discuss
a spectrum of a whole hypernucleus, which is intimately related to
the $\gamma$-ray spectroscopy, the measurement of which has been done with a germanium detector array (Hyperball) for $p$-shell hypernuclei~\cite{Hashimoto2006}.
To describe the energy spectrum of a whole Lambda hypernucleus based on a covariant energy density functional, one needs to treat properly the coupling between the unpaired valence hyperon and nuclear core excitations. To this end, one can construct the wave function for single-$\Lambda$ hypernuclei with an even-even nuclear core as follows,
 \begin{equation}
 \label{wavefunction}
 \displaystyle \Psi_{IM}(\vec{r}_{\Lambda},\{\vec{r}_N\})
 =\sum_{j\ell I_c}  {\mathscr R}_{j\ell I_{c}}(r_{\Lambda}) {\mathscr F}^{IM}_{j\ell I_c}(\hat{\vec{r}}_{\Lambda}, \{\vec{r}_N\}),
\end{equation}
where
\begin{equation}
 {\mathscr F}^{IM}_{j\ell I_c}(\hat{\vec{r}}_{\Lambda}, \{\vec{r}_N\})
= [{\mathscr Y}_{j\ell}(\hat{\vec{r}}_{\Lambda})\otimes
\Phi_{I_c}(\{\vec{r}_N\})]^{(IM)}
\end{equation}
with $\vec{r}_{\Lambda}$ and $\vec{r}_N$
being the coordinates for the $\Lambda$ hyperon and the
nucleons, respectively.
In this equation,
$I$ is the total angular momentum and $M$ is its projection onto the
$z$-axis for the whole $\Lambda$ hypernucleus.
${\mathscr R}_{j\ell I_{c}}(r_{\Lambda})$
and ${\mathscr Y}_{j\ell}(\hat{\vec{r}}_{\Lambda})$ are
the four-component radial wave function and
the spin-angular wave function for the $\Lambda$ hyperon, respectively.

The wave function of the nuclear core part, $\Phi_{I_cM_c}(\{\vec{r}_N\})$,
is given as a superposition of particle-number and angular-momentum
projected RMF+BCS states, $\vert \varphi(\beta)\rangle$, that is,
 \begin{equation}
 \vert \Phi_{I_cM_c}\rangle
 =\sum_\beta f_{I_c NZ}(\beta)
\hat P^{I_c}_{M_cK} \hat P^N\hat P^Z\vert \varphi(\beta)\rangle,
\label{GCM}
 \end{equation}
 where $\hat P^{I_c}_{M_cK}$, $\hat P^N$, $\hat P^Z$ are
the projection operators onto good numbers of angular momentum,
neutrons and protons, respectively. The  mean-field wave functions
$\vert \varphi(\beta)\rangle$ are a set of Slater determinants of
quasi-particle states with different quadrupole deformation $\beta$.
For simplicity, we consider only the axial deformation for the nuclear
core and thus the $K$ quantum number is zero in Eq. (\ref{GCM}).
The weight factor $f_{I_c NZ}(\beta)$ is determined
by solving the Hill-Wheeler-Griffin equation.
%We call this scheme
%a generator coordinate method (GCM) plus particle-number (PN) and one-dimensional %angular-momentum (1DAM)
%projections, GCM+PN1DAMP.

\begin{figure}[t]
  \centering
 \includegraphics[width=11cm]{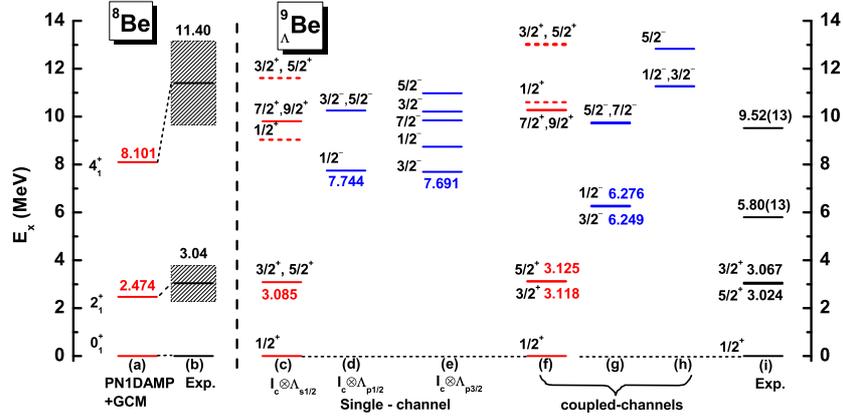}
 \caption{The low-energy excitation spectra of $^{8}$Be (the columns (a) and (b)) and  $^{9}_\Lambda$Be. For $^{8}$Be, the GCM+PN1DAMP calculations shown in the column (a) are compared with the experimental data. For $^{9}_\Lambda$Be, the columns (c), (d), and (e) show the results of the single-channel calculations for the $\Lambda$ particle in $s_{1/2}$, $p_{1/2}$, and $p_{3/2}$ channels, respectively. The columns (f), (g), and (h) show the results of the coupled-channels equations, which are compared with the experimental data~\cite{Hashimoto2006} shown in the column (i). This figure is taken from Ref.~\cite{MHYM14}.}
  \label{Be9L:spectra}
\end{figure}

Substituting Eq. (\ref{wavefunction}) to the Dirac equation for the
whole hypernucleus, $H|\Psi_{IM}\rangle = E_I|\Psi_{IM}\rangle$, where $H$ is the relativistic Hamiltonian derived from the effective Lagrangian density for the whole $\Lambda$ hypernucleus, one can derive the coupled-channels equations for ${\mathscr R}_{j\ell I_{c}}(r_{\Lambda})$, in which the coupling potentials are given in terms of the transition densities.

Figure~\ref{Be9L:spectra} displays the low-energy excitation spectra of $^{8}$Be and  $^{9}_\Lambda$Be, in comparison with the available data~\cite{MHYM14}.
For the $N\Lambda$ effective interaction term, the vector and scalar types of four-fermion coupling terms in Ref.~\cite{Tanimura2012Phys.Rev.C14306} are adopted. The coupling strengths are fitted to the $\Lambda$ binding energy.
One can see that a good agreement with the data is obtained for both $^{8}$Be and  $^{9}_\Lambda$Be.  The low-lying states of $^{9}_\Lambda$Be are categorized into
 three rotational bands, whose structures are confirmed by the calculated $B(E2)$ relations.
It is remarkable that the present calculation reconfirms an interesting prediction of the cluster model
that the strong coupling of a hyperon to the collective rotation is realized when
the $\Lambda$ is in the $p$-orbit \cite{MBI83}. Among these rotational bands,
the column (g) corresponds to what they called genuine hypernuclear states,
which are also referred to as the supersymmetric states having the SU3 symmetry $(\lambda\mu)=(50)$
of $s^4p^5$ shell-model configuration. These states do not have corresponding states in the
ordinary nucleus, $^9$Be, because of the Pauli principle of the valence neutron.
Moreover, according to our calculations, the measured state in $^{9}_\Lambda$Be with excitation energy of 5.80(13) MeV is actually a mixture of two negative-parity states with $J^\pi=3/2^-$ and $1/2^-$.

In view of the success of this novel method for the low-energy spectrum of Lambda hypernuclei, it would be interesting to extend this method to study hypernuclear magnetic moments and $M1$ transitions in the low-lying states. In particular, the determination of $\Lambda N$ interaction using the spectroscopic data of hypernuclei based on this method
is an interesting future work.

\section{Anti-Lambda hypernuclei}

\subsection{Polarization effect of anti-Lambda in nucleus}

Antibaryons are interesting particles for nuclear physics. They are building blocks of antimatter that can be  produced in the laboratory~\cite{Martin2013}. In free space, baryons and antibaryons promptly annihilate each other when  they close contact. Attempts have been made to find $\bar NN$ bound states with mass close to the threshold~\cite{Bai2003PhysRevLett.91.022001}. In a nuclear medium, many interesting predictions concerning the antibaryon behavior have been made. In particular, the appearance of antinucleon or antihyperon bound states in nuclei is one of the most popular  conjectures~\cite{Auerbach1986,Mao1999,Burvenich2002PLB,Mishustin2005PhysRevC.71.035201,Larionov2008PhysRevC.78.014604,Chen2010,Song2010}. It is demonstrated that the presence of a real antibaryon leads to a drastic rearrangement of a target nucleus, resulting in a significant increase of its binding energy and local compression.

In the RMF approach with meson-exchange for anti-$\Lambda$ hypernuclei, the coupling strengthes of $\bar\Lambda$ and mesons are related to those of $\Lambda$ and mesons according to the G-parity transformation. To take into account possible deviations from the G-parity symmetry in a many-body system, one can introduce an overall scaling $0\leq\xi\leq1$~\cite{Mishustin2005PhysRevC.71.035201}, namely,
 \beq
 g_{\sigma \bar\Lambda} = \xi  g_{\sigma\Lambda},~~ g_{\omega \bar\Lambda} = -\xi  g_{\omega\Lambda}.
\eeq

 \begin{figure}[t]
 \centering
  \includegraphics[width=8cm]{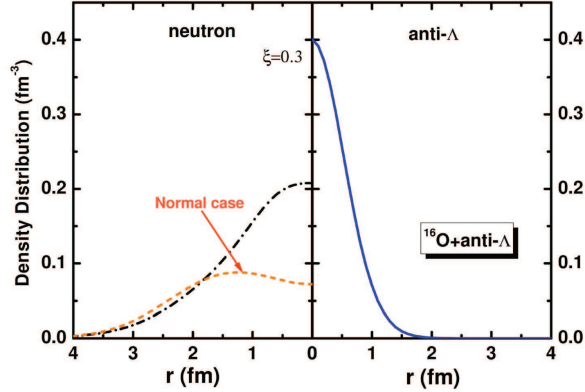}
   \caption{Density distribution for neutrons (left panel) and $\bar \Lambda$ (right panel) in $^{16}$O$+\bar\Lambda$ system, where the $\bar\Lambda$ occupies the lowest $s$ orbit. The results are given by the RMF calculations with the PK1 force for the effective $NN$ interaction and $g_{\sigma\Lambda}/ g_{\sigma N}=g_{\omega\Lambda}/ g_{\omega N}=2/3$ (based on the SU(3) flavor symmetry), $\xi=0.3$ for the $\bar \Lambda N$ interaction. The density distribution for neutron in $^{16}$O is plotted for comparison. Taken from Ref.~\cite{Song2010}.}
   \label{density:antilambda}
 \end{figure}
 \begin{figure}[]
 \centering
  \includegraphics[width=9cm]{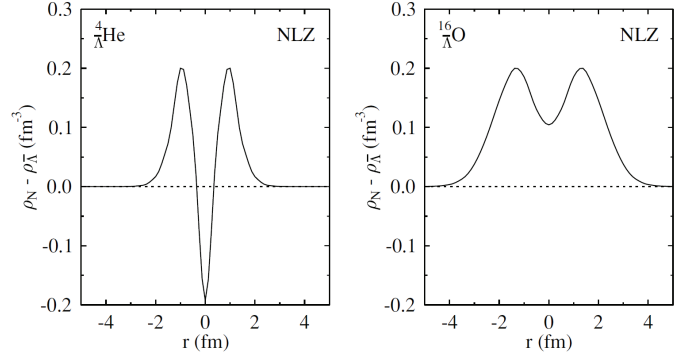}
   \caption{Distribution of the net baryon density ($\rho_N-\rho_{\bar\Lambda}$) in the $^4$He$+\bar\Lambda$ (left)  and $^{16}$O$+\bar\Lambda$ (right) systems calculated within the
RMF model with the NLZ~\cite{Rufa88PhysRevC.38.390} parameter set. Taken from Ref.~\cite{Mishustin2005PhysRevC.71.035201}.}
   \label{density:antilambda2}
 \end{figure}

The polarization effect due to the $\bar \Lambda$ on the nuclear density distribution of $^{16}$O is illustrated in Fig.~\ref{density:antilambda}. It is seen that the presence of $\bar \Lambda$ compresses the protons and neutrons into the center of the nucleus with the central nuclear density (the sum of densities for proton and neutron) up to $2-3$ times of saturation density. The distribution of $\bar \Lambda$ indicates its interaction range, which is in between $1-2$ fm. It means that only the nucleons close to nuclear center will be strongly influenced by the $\bar \Lambda$. The distribution of the net baryon density ($\rho_N-\rho_{\bar\Lambda}$) in the $^4$He$+\bar\Lambda$  and $^{16}$O$+\bar\Lambda$ systems from the RMF model calculation
with the NLZ~\cite{Rufa88PhysRevC.38.390} parameter set is shown in Fig.~\ref{density:antilambda2}. Because the $\bar \Lambda$ has $-1$ baryon number, it depresses the density of baryon numbers in the center, even to an negative value in $^4$He$+\bar\Lambda$ due to the strong localization of $\bar\Lambda$~\cite{Mishustin2005PhysRevC.71.035201} as also shown in Fig.~\ref{density:antilambda}.

\begin{figure}[]
 \centering
  \includegraphics[width=9cm]{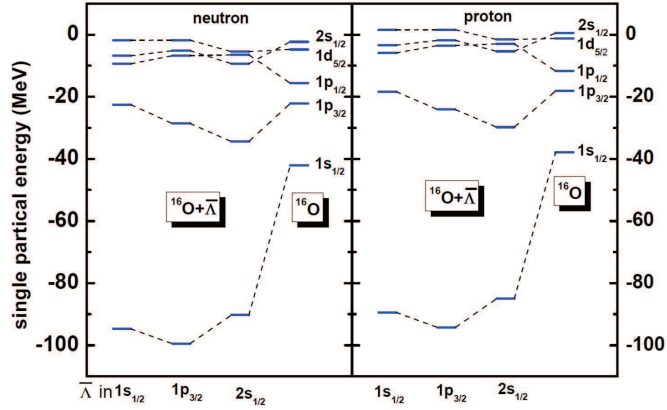}
   \caption{The single-particle energy of neutron and proton in $^{16}$O$+\bar\Lambda$ system, where the $\bar\Lambda$ occupies $1s_{1/2}, 1p_{3/2}$ and $2s_{1/2}$, respectively. The single-particle energy in $^{16}$O (last column) is plotted for comparison.}
   \label{spe:antilambda}
 \end{figure}

The nucleons in $1s_{1/2}$ state distribute mainly around the nuclear center and thus feel a deeper potential due to the presence of the $\bar\Lambda$, which results in a larger binding energy, as seen in Fig.~\ref{spe:antilambda}. However, the nucleons at higher energy states are far away from the center and therefore are not much influenced by the $\bar\Lambda$, except the changes in spin-orbit splitting.
That is, the polarization effect of anti-Lambda on single-particle structure of atomic nuclei
varies depending on which state it occupies. For nucleons in 1s1/2 state, the anti-lambda makes
it deepest bound when it occupies 1p3/2 state.
For the nucleons in 1p3/2 state, on the other hand,
the largest polarization effect is found when the anti-Lambda occupies 2s1/2 state.
This phenomenon can also be understood from the range of the polarization effect by
$\bar \Lambda$.

\subsection{Spin symmetry in the spectrum of anti-Lambda hyperon}

\begin{figure}[]
\includegraphics[width=12cm]{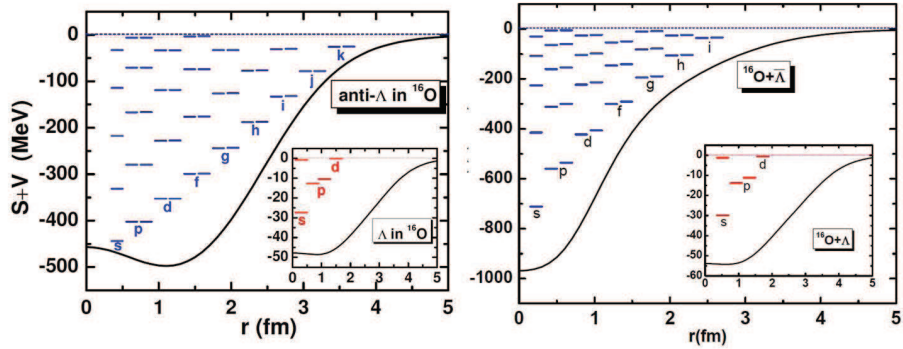}
\caption{Mean-field potential and single-particle energy spectrum of $\bar{\Lambda}$ in
$^{16}$O (the left panel) and
those of $\bar{\Lambda}$ in $^{16}$O$+\bar\Lambda$ system (the right panel).
The inset gives the potential and spectrum of $\Lambda$ in $^{16}$O and that of $\Lambda$ in $^{16}$O$+\bar\Lambda$ system. The polarization effect of $\bar{\Lambda}$ or $\Lambda$ in $^{16}$O is taken into account in the right panel. $\xi=1$ is used for the $\bar\Lambda N$ interaction. Taken from Ref.~\cite{Song2009}.} \label{fig:spectrum1}
\end{figure}

\begin{figure}[t]
\centering
\includegraphics[width=8cm]{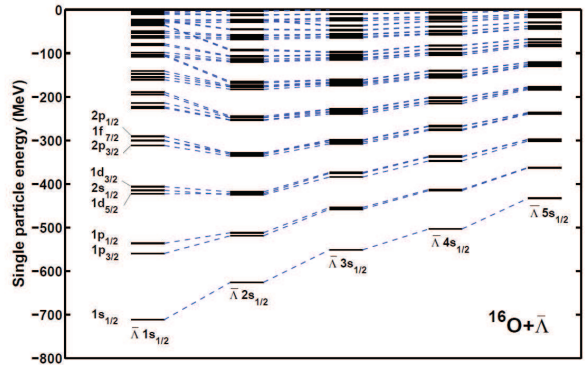}
\caption{Single-particle energy spectrum of $\bar{\Lambda}$ in $^{16}$O$+\bar\Lambda$ system, where  the $\bar\Lambda$ occupies different $s$ orbit, respectively. } \label{fig:spectrum2}
\end{figure}

The potential and single $\bar{\Lambda}$ spectrum in $^{16}$O and those of $\bar{\Lambda}$ in $^{16}$O$+\bar\Lambda$ system are plotted in Fig.~\ref{fig:spectrum1}.  As seen in the left panel of Fig.~\ref{fig:spectrum1}, when the rearrangement effect of $\bar{\Lambda}$ or $\Lambda$ is not taken into account in $^{16}$O, the energies of $\bar\Lambda$ spin-orbit doublets are almost identical. The energy differences $\epsilon_{\bar{\Lambda}(nl_{j=l-1/2})}-\epsilon_{\bar{\Lambda}(nl_{j=l+1/2})}$
in the anti-Lambda spectrum are actually around 0.09-0.17 MeV for $p$ states, which are much smaller than that in $\Lambda$ spectrum, 2.26 MeV, estimated without the tensor
coupling. It has been discussed in ordinary atomic nuclei that there is good pseudo-spin symmetry in single-nucleon spectrum and spin symmetry in the spectrum of anti-nucleon~\cite{Zhou2003}. However, if the polarization effect of $\bar{\Lambda}$ on the mean-field is taken into account, the spin symmetry will be destroyed as illustrated in the right panel of Fig.~\ref{fig:spectrum1}. With the increase of the energy of the occupied $\bar{\Lambda}$ state, the polarization effect of $\bar{\Lambda}$
decreases, and thus the spin symmetry in the single-particle energy spectrum of $\bar{\Lambda}$ is recovered gradually as exhibited in  Fig.~\ref{fig:spectrum2}.

A further study on the spin symmetry in the single-particle energy spectrum of $\bar{\Lambda}$ was made in Ref.~\cite{Song2011}. It has been found that the tensor coupling enlarges the spin-orbit splittings for $\bar\Lambda$ by a factor of 5 but has a negligible effect on the wave functions of $\bar\Lambda$. This indicates that the spin symmetry in anti-$\Lambda$-nucleus systems is still a good approximation. 

\section{Multi-strange systems}
\subsection{Multi-{$\Lambda$} hypernuclei}

So far, we have focused on the properties of single-$\Lambda$ hypernuclei.
In this section, we will discuss properties of multi-$\Lambda$ hypernuclei.
Among multi-$\Lambda$ hypernuclei, so far only double-$\Lambda$ hypernuclei have been
experimentally found in an emulsion ~\cite{Takahashi01,Ahn13}.
The mass of double-$\Lambda$ hypernuclei is often discussed in terms of
the $\Lambda-\Lambda$ interaction energy defined as
\begin{equation}
\Delta B_{\Lambda\Lambda}(^{~A}_{\Lambda\Lambda}Z)\equiv B_{\Lambda\Lambda}(^{~A}_{\Lambda\Lambda}Z)
-2B_\Lambda(^{A-1}_{~~\Lambda}Z),
\end{equation}
where $B_{\Lambda\Lambda}$ and $B_\Lambda$ are the two-$\Lambda$ and one-$\Lambda$
binding energies, respectively. The experimental value of $\Delta B_{\Lambda\Lambda}$
for the $^{~6}_{\Lambda\Lambda}$He hypernucleus is  0.67$\pm$0.17 MeV ~\cite{Ahn13}, which
indicates that the $\Lambda\Lambda$ interaction is attractive but weak.

\begin{figure}[tb]
%\centering
\begin{center}
\includegraphics[width=6cm]{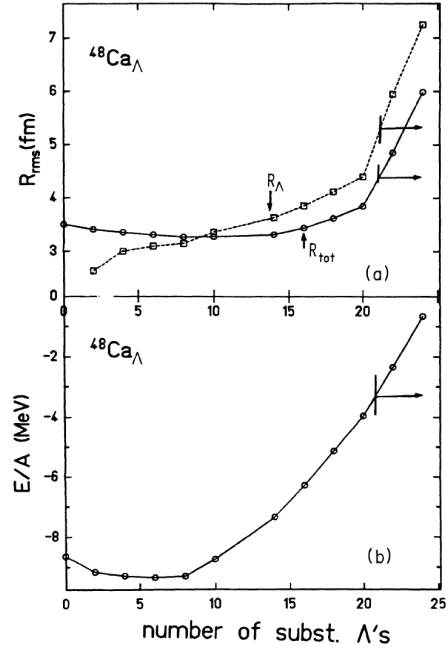}
\caption{The root mean square radii (the upper panel) and the binding energies per baryon
(the lower panel) as a function of the number $n$ of $\Lambda$ particle  in $^{48}_{n\Lambda}$Ca hypernuclei. The mass number of each hypernucleus is kept to be 48 by changing the neutron number according to the number of $n$. The arrows indicate the particle unbound region. Taken from Ref. ~\cite{RSM90}.
}
\label{fig-rufa}
\end{center}
\end{figure}

For the meson exchange version of relativistic mean-field theory, it is straightforward to extend it
from single-$\Lambda$ to multi-$\Lambda$ systems~\cite{SYT06}.
That is, only the $\Lambda$ density in the Klein-Gordon
equations is changed and the structure of the Dirac equation remains the same. Since the $\Lambda\Lambda$
interaction is weak, a residual $\Lambda\Lambda$ pairing interaction is usually neglected.
This simply leads to a larger impurity effect of $\Lambda$ particle ~\cite{SWHS10,MH12}.
Fig. \ref{fig-rufa} shows the root-mean-square radii and the total energy per baryon as a function of
the number of $\Lambda$ particle, $n$, in $^{48}_{n\Lambda}$Ca hypernuclei ~\cite{RSM90}.
Notice that the neutron number is varied for each $n$ so that the total mass number is kept to be 48.
The energy and the radii show non-monotonic behaviors. The $\Lambda$ particles first increase the binding
energy and the radius decreases as its consequence. When the $\Lambda$ particle start filling weakly bound
single-particle $\Lambda$ levels, the binding then becomes weaker and the radius increases.
This behavior is understandable since there is no Pauli principle between the $\Lambda$ particles and the nucleons.
A similar conclusion has been reached also in Refs. ~\cite{MZ89,MZ93} and in a very recent study~\cite{Ikram2014}.

\subsection{Neutron stars}

We next discuss neutron stars, which offer natural laboratories to explore the equation of
state (EOS) at low temperature and high densities ~\cite{HH00,LP07,Lattimer12,SR04}.
In particular, the observed masses of neutron stars have provided a strong constraint
on the EOS. Many neutron starts have been observed to have a mass of
around 1.4 $M_\odot$~\cite{LP07}, where $M_\odot$ is the solar mass, and this value has been
regarded as a standard value for many years.

\begin{figure}[tb]
%\centering
\begin{center}
\includegraphics[width=10cm]{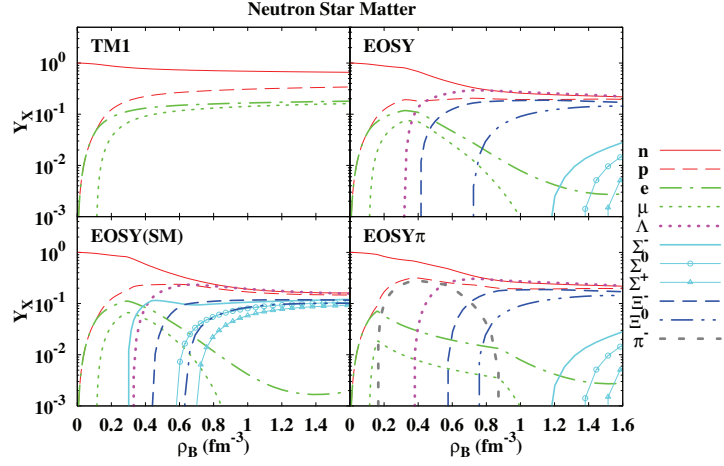}
\caption{
The fraction of barions and leptons in neutron star matter obtained with the
relativistic mean field calculations. The upper left panel (TM1) shows the result without
hyperons, while the right upper panel (EOSY) and the left lower panel (EOSY(SM)) show
the results with hyperons, with a repulsive and an attractive $\Sigma$ potentials, respectively.
The right lower panel (EOSY$\pi$) shows the result with pion contribution.
Taken from Ref. ~\cite{Ishizuka08}.
}
\label{fig-neutronstar}
\end{center}
\end{figure}

In addition to the non-relativistic approaches ~\cite{SBLCL98,BBS98,Vidana00,SPRV06,NYT02,TNYT06},
the relativistic approach has also been applied both to
strange matter ~\cite{RSM90,BLMM91} and to neutron star matter with hyperons
~\cite{GM91,ST94,KPE95,SM96,Long12,Shen02,ShenEOS,Ishizuka08,Tsubakihara10,TO13}.
One of the most important conclusions of these studies is that hyperons should admix
in dense neutron star matter (see Fig. \ref{fig-neutronstar}), in which the first hyperon
to appear is either $\Lambda$ or $\Sigma^-$ depending on the properties of single-particle
potential for $\Sigma$ particle.
The hyperon admixture can be easily understood using the non-relativistic
Thomas-Fermi approximation (see {\it e.g.,} Refs.~\cite{Walecka2008,MRS2010} for the
Thomas-Fermi approximation).
In the Thomas-Fermi approximation, the total energy (including the mass energies) for
matter with $N_n$ neutrons and $N_\Lambda$ $\Lambda$ particles is given by
\begin{equation}
E=m_Nc^2\cdot N_n+\frac{3}{5}N_nE_{Fn}+m_\Lambda c^2\cdot N_\Lambda+\frac{3}{5}N_\Lambda
E_{F\Lambda},
\end{equation}
where $E_{Fn}$ and $E_{F\Lambda}$ are the Fermi energy for the neutrons and the $\Lambda$ particles, respectively, and $m_N$ is the nucleon mass.
For $N_n=(1-x)N_B$ and $N_\Lambda=xN_B$, the total energy per baryon then reads
\begin{equation}
\frac{E}{N_B}=m_Nc^2+\frac{3}{5}E_{Fn}+\left[(m_\Lambda-m_N)c^2+\frac{3}{5}(E_{F\Lambda}-E_{Fn})\right]x,
\end{equation}
with
\begin{eqnarray}
E_{Fn}&=&\frac{\hbar^2}{2m_N}\left(3\pi^2\cdot(1-x)\rho_B\right)^{2/3}, \\
E_{F\Lambda}&=&\frac{\hbar^2}{2m_\Lambda}\left(3\pi\cdot x\rho_B\right)^{2/3},
\end{eqnarray}
where $\rho_B$ is the baryon density.
This equation indicates that the total energy is minimized with $x$=0 (and thus $E_{F\Lambda}$=0)
when the neutron
Fermi energy $E_{Fn}$ is smaller than $5/3\cdot(m_\Lambda-m_N)c^2$ while $x$ becomes a finite
value after $E_{Fn}$ exceeds $5/3\cdot(m_\Lambda-m_N)c^2$.
This result is a physical consequence of the fact that
the increase of neutron energy due to the Pauli principle can be avoided by converting a
fraction of neutrons to $\Lambda$ particles, although
the actual value of the critical density for the hyperon admixture depends on
details of baryon-baryon interactions.

\begin{figure}[tb]
%\centering
\begin{center}
\includegraphics[width=6cm]{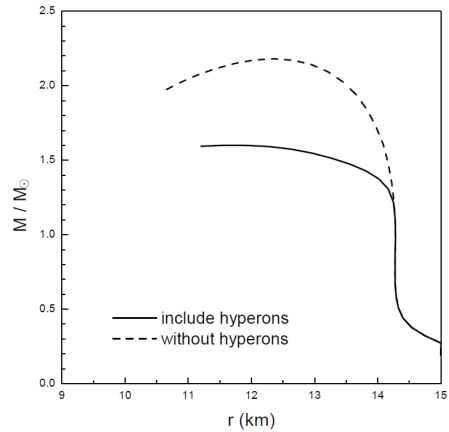}
\caption{
A mass-radius relation for neutron stars obtained with
RMF. Taken from Ref. ~\cite{Shen02}.}
\label{fig-nsmass}
\end{center}
\end{figure}

Another important conclusion on the previous theoretical studies is that the EOS is softened when
the admixture of hyperons is taken into account. This results in a smaller maximum mass of
neutron stars. Fig. \ref{fig-nsmass} shows a mass-radius relation obtained by solving the
Tolman-Oppenheimer-Volkov equation ~\cite{SR04} with the EOS of Shen {\it et al.}
~\cite{ShenEOS,Shen02}, which was constructed based on RMF. One can see that the maximum mass
of neutron stars is around 2.2 $M_\odot$ when the hyperon admixture is not considered whereas
it decreases to 1.6 $M\odot$ when the hyperon admixture is taken into account.
Very recently, a large neutron mass of (1.97$\pm$ 0.04)$M_\odot$ was observed
using the Shapiro delay for the binary millisecond pulsar J1614-2230 ~\cite{DPR11}.
Most of EOS for hyperonic neutron star matter is not compatible with such a large neutron star
mass, and it has remained a big open problem in the present nuclear physics.

More detailed discussions on neutron stars will be given in Chap. 15.

\section{Summary}

Hypernuclei with one or more hyperons have provided interesting 
and unique opportunities to study nuclear many-body problems, in addition to the main 
motivation of hypernuclear physics, that is, to extract the information on nucleon-hyperon and 
hyperon-hyperon interactions. 
One of the important subjects along this direction is to clarify the impurity effect 
of $\Lambda$ hyperons, that is, to clarify how several properties, such as deformation and 
collective excitations, of atomic nuclei are 
influenced by adding a $\Lambda$ particle. 
A relativistic mean-field approach has been one of the standard tools for this purpose. 
An important point is that the spin-orbit splitting is described naturally with the 
relativistic approach. It has been known experimentally that the spin-orbit splittings 
for $\Lambda$ particles are much smaller than that for nucleons, and the relativistic approach 
has played an important role in hypernuclear physics, especially at the early stage of 
research in the late 1970's. 

The relativistic approach has predicted an interesting impurity effect on nuclear deformation. 
That is, there are certain nuclei for which the deformation disappears when a $\Lambda$ 
particle is added to them. This phenomenon is difficult to see with non-relativistic approaches 
since the polarization effect of $\Lambda$ particle is weaker than the relativistic approach. 
 
A drawback of the mean-field approach is that
the pure mean-field approximation does not
yield a spectrum of hypernuclei due to the broken symmetries (this applies both to 
the non-relativistic and the relativistic approaches).
In order to cure this problem, the so called beyond mean-field approach has been 
developed for ordinary nuclei (see Chap. 10 in this Book). 
The beyond mean-field approach includes 
a restoration of the broken symmetries by angular momentum
and particle number projections.
It also includes the shape fluctuation effect, which is 
taken into account with the generator
coordinate method (GCM) or its approximation, the collective
Hamiltonian approach. 
The beyond mean-field approach has been applied to hypernuclei 
and the spectra of hypernuclei have now been constructed based on (Covariant) Density 
Functional Theory, using the microscopic particle-rotor model. 
A nice feature of hypernuclei is that the problem becomes simpler 
because of the absence of Pauli principle between a $\Lambda$ particle and nucleons in a 
core nucleus. 

Another important subject of hypernuclear physics is neutron star. 
There is almost no doubt that hyperons appear in dense neutron star matter, since 
it is a natural consequence of Pauli principle. 
A problem is that the hyperon admixture softens the equation of state and decreases 
the maximum neutron star mass. In fact, most EOS for hyperonic neutron star matter are 
incompatible with the recently observed large neutron star mass ($\sim 2M_\odot$), and 
it has remained a big challenge in the nuclear physics community to account for the 
large neutron mass together with the hyperon admixture. 
Hopefully, new experimental investigations of hypernuclei at
the next generation
experimental facilities, {\it e.g.,} the J-PARC facility, and theoretical studies 
with such new data will resolve this big problem in near future. 

\bigskip

\section*{Acknowledgments}
We thank T. Koike, Z. P. Li, H. F. L\"{u}, H. Mei, J. Meng,  T. Motoba, H. Sagawa, H.-J. Schulze, C. Y. Song, H. Tamura, Y. Tanimura, M. T. Win, W. X. Xue for useful discussions and their contributions to the work reviewed in this chapter. This work was supported in part by the Tohoku University Focused Research Project ``Understanding the origins for matters in universe", JSPS KAKENHI Grant Numbers 25105503 and 26400263, the National Natural Science Foundation of China under Grant Nos. 11305134, 11105111, and the Fundamental Research Funds for the Central Universities (XDJK2010B007 and XDJK2013C028).

%\include{ref}
 % for BibTeX users
\bibliographystyle{ws-book-har}    % Bibliography: Author-Date system
\bibliography{ch7}

\end{document}